# Nodal structures of few electron atoms

S Datta [1,2],

[1] Department of Physics, Icfai Foundation for Higher Education , Hyderabad- 501203, India

J M Rejcek [2]   J. L. Fry [2],

[2] Department of Physics, University of Texas at Arlington, Arlington, Texas 76019

In this paper the nodal structure for He atom is revisited and a procedure for identifying the symmetry properties and nodal structure of the time and spin independent Schrödinger equation will be presented. Application are made to the few electron atoms, and explicit, new nodal surface equations given for states of various symmetries. In the construction of trial functions for variational calculations in quantum mechanics, and in the choice of nodal surfaces needed in Monte Carlo calculations to obtain fermion solutions to Schrodinger's equation for spin independent Hamiltonians, the accuracy of the calculations can be strongly influenced by the initial choice of the trial functions. We incorporated these newly found nodal conditions in the trial functions and energy eigenvalues were calculated with a better accuracy for different symmetries.

1. **Inroduction:**

Quantum Monte Carlo methods(QMC) are widely used techniques for the calculation of the lowest energy state corresponding to a particular symmetry of the quantum many body system . Despite their long history dating back to early 1940's QMC have suffered from several difficulties. The first two difficulties can be attributed to the lack of mathematical justification and unreasonably high computing time resources. The third and most severe difficulty these methods encounter is the so called sign problem for the fermionic systems, which results in the numerical instability and lack of convergence due to the cancellation of the positive and negative terms in the eigenfunction expansions. The purpose of this paper is to apply the Generalized Feynman-Kac (GFK) method to atomic systems in a completely new way and investigate the nodal structure of simple atomic systems. For to calculate the energy eigenvalue for a quantum many body system with GFK formalism, we need a trial function and the choice of which strongly determines the accuracy of the calculation. So as to obtain faster convergence and good accuracy, apart from satisfying the usual quantum mechanical properties ( $\psi$ must be finite, single-valued, continuous at the point of the configuration space ) the trial functions need to satisfy the following conditions:

(1) Cusp conditions
(2) Nodal conditions

In this paper we focus on the symmetry properties of the wave functions. The reasons that we investigate the symmetry properties of the wave functions are twofold. The first is to guarantee that we calculate the lowest energy corresponding to a symmetry of a particular representation. According to Simon the lowest energy with $\psi(\vec{r}) = 0$ on the boundary D (any region of the phase space) will have no additional nodes in D. $\psi(\vec{r})$ has to be strictly positive or negative in the interior of D.

Consequently, the $\psi(\vec{r})$ that changes sign must have higher energy. As a result, solution corresponding to a given energy, the trial function must have correct nodal surface which partitions the entire phase space into regions where $\psi(\vec{r})$ has constant sign.

In d-dimension, the condition $\psi(\vec{r}_1, \vec{r}_2....\vec{r}_n) = 0$ defines a $dn - 1$ dimensional nodal surfaces whereas corresponding to antisymmetry, there will be $dn - d$ dimensional surfaces. For one dimensional problems, these conditions coincide and antisymmetry alone specifies the full nodal surface.. But for higher dimensional problems antisymmetry alone cannot specify the full nodal surface. Hence for $d > 1$, antisymmetry alone cannot specify a part of the irreducible portion of the configuration space where the wave function has a constant sign. To get the full nodal surface one needs to consider symmetries other than permutation symmetry ( for example rotational symmetry of the many particle system).

Secondly, to obtain numerical solution of a many body problem efficiently, one needs to seek out some procedures which maximize the speed of convergence and minimize the computing time. The symmetry which we exploit via group theory is, in most cases, the symmetry of the Hamiltonian operator. Before discussing the symmetry of the atomic systems let us discuss a few ideas from group theory.

In this paper, in section 2 .1 , we review the preliminary ideas about group theory. In section 3 we deduce the nodal conditions associated with different symmetries. In section 4 we present the analytical forms of nodal structures and the energy eigenvalues using trial functions employing those nodal conditions. Finally we put our concluding remark in section 5.

# 2
## 2.1 The Group of the Schrödinger Equation for the many Particle Quantum Systems.

Before we deduce the actual nodal conditions, we need to outline the symmetry properties of a quantum mechanical many body system. Let us first consider a typical Hamiltonian for a quantum system of one particle of unit mass in atomic units:

$$H(\vec{r}) = -\frac{1}{2}(\frac{\partial^2}{\partial x^2} + \frac{\partial^2}{\partial y^2} + \frac{\partial^2}{\partial z^2}) + V(\vec{r})$$

**Theorem 1**

The set of coordinate transformations that leaves the Hamiltonian invariant form a group. The group is usually called the group of *Schrödinger Equation.*

For the Hydrogen atom or any other single particle system in which $V(\vec{r})$ is a function of $\vec{r}$ only, the group of the Schrödinger Equation is the group of all pure rotations in $R^3$. Now if we turn our attention to the invariant group of the Hamiltonian for an atom with N electrons :

$$H = -\frac{1}{2}\sum_i \Delta_i + \frac{1}{2}V_{ij} - \sum_j^N \frac{1}{|\vec{r}_i|} \qquad (2.1)$$

where $V_{ij}$ is the Coulomb interaction:

$$V_{ij} = \frac{1}{|\vec{r}_i - \vec{r}_j|}$$

This Hamiltonian remains unchanged by permutations of electron coordinates and by simultaneous rotation of all the electrons in $R^3$ as follows:

$$\alpha\vec{r}_i = \vec{r}_i, \quad |\alpha\vec{r}_i - \alpha\vec{r}_j| = |\vec{r}_i - \vec{r}_j| \text{ and } |\alpha\vec{r}_i| = |\vec{r}_i|$$

The kinetic energy term will be unaffected by permutation and rotation. The set of simultaneous rotations form a group, but it is not the group $\vartheta(3)$ or $\vartheta(3N)$, the group of all rotations in $R^3$ and $R^{3N}$ respectively. Its irreducible representations are needed to study the minimal nodal structure of many particle wave function which is the solution to the
Schrödinger Equation

$$H\psi(x_1,......x_n) = E\psi(x_1,......x_n)$$

The group structure may be deduced from $\vartheta(3N)$ and it turns out to be isomorphic to $\vartheta(3)$. Let us now consider the group $\vartheta(3N)$, group of all rotations in 3N dimensional space which preserves the length of the $3N$ vectors.

$$|\alpha_{3N}\vec{r}_{3N}| = |\vec{r}_{3N}|$$

Now restrict the rotation matrix to the subset, that is the subgroup of $\vartheta(3N)$.

$$\begin{pmatrix} \alpha_1 & 0........ & 0 \\ 0 & \alpha_2....... & 0 \\ 0 & 0 & \alpha_N \end{pmatrix} \equiv G_s$$

i.e., the rotations $\alpha_1 .......... \alpha_N$ are the 3 dimensional rotation matrices, so that only coordinates of each particle are rotated separately. Now consider the N set of rotations

$$\begin{pmatrix} \alpha_1 & 0........ & 0 \\ 0 & 1...... & 0 \\ 0 & 0 & 1 \end{pmatrix}, \begin{pmatrix} 1 & 0........ & 0 \\ 0 & \alpha_2...... & 0 \\ 0 & 0 & 1 \end{pmatrix} ........................ \begin{pmatrix} 1 & 0........ & 0 \\ 0 & 1....... & 0 \\ 0 & 0 & \alpha_N \end{pmatrix}$$

Each of these sets form a subgroup of $G_s$ and each is isomorphic to $\vartheta(3)$. The subgroups are invariant subgroup of $G_s \equiv \vartheta_s(3N)$. So $G_s$ is a direct product

$$G_s^{T_1,T_2,......T_n} = G_1(T_1) \otimes G_1(T_2) \otimes ..........G_N(T_N)$$

Thus for two particles, e.g.:

$$\Gamma_{(T_1T_2)}^{pq} = \Gamma_1^p(T_1) \otimes \Gamma_2^q(T_2)$$

$\Gamma_1, \Gamma_2$ are irreducible rep of $\vartheta(3)$

So all the irreducible of $G_s$ are known, since for a direct product group this equation provides irred rep of $G_s$. Actually we derive the irred rep of another group,

$$\begin{pmatrix} \alpha & 0....... & 0 \\ 0 & \alpha....... & 0 \\ 0 & 0....... & \alpha \end{pmatrix}$$

This is a subgroup of $G_s$ and is known as diagonal subgroup. It can be written as along (with $T_1 = T_2$ etc):

$$G_d^{(TT.........T)} = G_1(T) \otimes G_2(T) \otimes ........G_N(T)$$

Unfortunately this is a red rep of $G_d$ This diagonal subgroup is isomorphic to

$G_1, G_2 .............. G_N$ and $\vartheta(3)$

## 2.2 Nodal Structures and Symmetry Properties of a Quantum Mechanical Wave Function

In this section we deduce the nodal conditions for different symmetry of few electron systems. These conditions can be used to identify the minimal nodal structures required for the jth row of the ith irreducible representation of the invariant group of the Hamiltonian. For a given irreducible representation these nodal conditions are necessary to determine the lowest eigenvalue and eigenfunction for that symmetry.

Elementary group theory may be employed to prove that the eigenfunction of the Schrödinger Equation for the construction of irreducible representations of the invariant group $\mathbf{G}$, of the Hamiltonian operator, $\mathbf{H}$. This is equivalent to the statement that the eigenfunctions must transform like one of the rows of the reps of $\mathbf{G}$. The group is defined as the set of operators which commute with the Hamiltonian operator.

$$\mathbf{G} = \left\{ \mathbf{O}_\mathbf{l} \big| [\mathbf{O}_\mathbf{l}, \mathbf{H}] \right\} = 0$$

Using the definition of projection operator we write

$$\mathbf{P}_\mathbf{j}^\mathbf{i} \psi = \psi_\mathbf{j}^\mathbf{i}$$

where $\psi_j^i$ is the part of $\psi$ which transforms like $j$th row of the $i$th rep of $\mathbf{G}$

$$\mathbf{P}_\mathbf{j}^\mathbf{i} \psi = 0$$

if no part of $\psi$ transforms like the jth row of ith rep. Since N dimensional reps correspond to an N-fold degeneracy of eigenvalues, an arbitrary linear combination of a given set of solutions transforming like a particular rep is also a solution with the same eigenvalue, i.e., by a unitary transformation we can generate N basis functions for an *equivalent* representation. The particular representation employed is a matter of convenience and can result in a set of functions with different symmetries. The corresponding projection operatorcan easily be constructed. For example, in molecular and solid state calculations real cubic harmonics are often employed instead of complex spherical harmonics.

In what follows the symmetry properties of a function transforming like $\psi_j^i$ are examined. If $\psi_j^i$ is a function belonging to the jth row of ith rep of $\mathbf{G}$ the result of one of the operations in $\mathbf{G}$ acting on the function f in general

$$\mathbf{O}_\mathbf{l} \psi_\mathbf{j}^\mathbf{i} = \sum_\mathbf{m} \mathbf{O}_{\mathbf{jm}}^\mathbf{l} \psi_\mathbf{j}^\mathbf{i}$$

It is useful to define a subset of operators of $\mathbf{G}$ by the condition

$$\mathbf{O}_\mathbf{l} \psi_\mathbf{j}^\mathbf{i} = \lambda_\mathbf{k} \psi_\mathbf{j}^\mathbf{i} \qquad (2.2)$$

Since the representations are unitary, the eigenvalues are unimodular. A necessary and sufficient condition that an operator in $\mathbf{G}$ belongs to this set is

$$[\mathbf{O}, \mathbf{P}_\mathbf{j}^\mathbf{i}] = \mathbf{0} \qquad (2.3)$$

This set of operations form a group, $\mathbf{G}_\mathbf{j}^\mathbf{i}$, which is a subgroup of $\mathbf{G}$.

$$\mathbf{G}_\mathbf{j}^\mathbf{i} = \left\{ \mathbf{O}_\mathbf{k} \big| \mathbf{O}_\mathbf{k}, \mathbf{P}_\mathbf{j}^\mathbf{i} \right\} = \mathbf{0}, \mathbf{O}_\mathbf{k} \subset \mathbf{G} \qquad (2.4)$$

Note that from Eq(2.3) knowledge of **G** is all that is necessary to find $\mathbf{G}_j^i$ (the function $\psi_j^i$ is not required). The group $\mathbf{G}_j^i$ is called the eigengroup for row j of the ith rep since it contains all the operators of **G** for which $\psi_j^i$ is an eigenfunction. If representations are constructed so that the real basis functions result, $\lambda = \pm 1$. For any operator with $\lambda = -1$, Eq(2.2) reads

$$\mathbf{O}_k \psi_j^i(\mathbf{r}) = \psi_j^i(\mathbf{O}_k^{-1}\mathbf{r}) = -\psi_j^i(\mathbf{r}) \tag{2.5}$$

so $\mathbf{O}_k^{-1}\mathbf{r}_0 = \mathbf{r}_0$ (2.6)

defines a node at $\mathbf{r}_0$ of any function transforming like jth row of the ith rep.

Operators belonging to $\mathbf{G}_j^i$ with $\lambda = +1$ leave the function unchanged. These operators form a subgroup $\mathbf{G}_{js}^i$ of $\mathbf{G}_j^i$ and **G**. It is the group of invariant operators for $\psi_j^i$. Given a function value at $\mathbf{r}_1$, this group of operations locates a set of points, $\mathbf{O}_k \mathbf{r}_1$ at which the function value is the same as it is at the point $\mathbf{r}_1$. This includes any of the nodal points, $\mathbf{r}_0$, located by the operators of $\mathbf{G}_j^i$ which have $\lambda = -1$, but also includes nodes of a particular $\psi_j^i$ which are not required by any of the operations of $\mathbf{G}_j^i$, but are the consequences of orthogonality or, e.g., the radial dependence of a function, or a symmetry node not revealed by Eq.(2.5):

$$\psi_j^i(\mathbf{r}_0) = 0 \tag{2.7}$$

Nodes given by Eq(2.7) occur in all the excited states of each rep. Eq(2.5) provides the necessary condition for all $\psi_j^i$, but for the lowest energy eigenfunction of each rep, it may also be a sufficient condition if it closes the space.

Let $D$ be any region in the phase space which is (semi) bounded by the nodal surfaces obtained by the application of Eq(2.5). A solution of Schrödinger's Equation inside $D$ with boundary condition $\psi(\mathbf{r}) = 0$ on the surface of $D$ can be mapped into all of phase space with the operations of $\mathbf{G}_j^i$, to obtain the eigenfunction throughout the phase space if desired.(although it is not necessary) Thus it satisfies everywhere Schrödinger's Equation, obtain the eigenfunction throughout the phase space along with its boundary of condtions, and it also has the required eigenfunction transformation properties of $\mathbf{G}_j^i$. By the result of Simon the lowest energy solution with $\psi(\mathbf{r}) = 0$ on the boundary of $D$ will have no other nodes in $D$. Since the solution found satisfies Schrödinger's Equation.the imposed boundary conditions, and the requirements of group theory, it is a sought after solution. All other solutions will have nodes inside $D$, and must have a higher energy. *If Eq (2.5) finds a set of nodal surfaces (if it closes the space) these provide a necessary and sufficient set of conditions for the lowest energy solution for each rep.* But in general this does not represent the equation of a closed surface. Actually in most cases the nodal conditions we get, are necessary but not sufficient. Recall, however, that the nodal conditions are not unique, but depend upon which of many possible equivalent representations have been employed to generate $\mathbf{G}_j^i$.. The nodal surfaces may be obscured in such a way that Eq (2.5) provided only nodal points or a null set of operators, but Eq(2.7) provides the other nodes. An example is the $l = 2, m = 0$ spherical harmonic function. In these instances another row, or an equivalent representation can provide the necessary information through Eq(2.5).

The simplest application Eq(2.5) is the hydrogen atom. Using cubic harmonic basis functions instead of spherical harmonics we construct representations with real basis functions, at the expense of the loss of good quantum number m, but preserving the quantum number l and the energy eigenvalue E.

Functions with desired m may be obtained after the calculations by appropriate transformation. The group $\mathbf{G}$ is the rotation group in three dimensions, $\mathbf{R^{3N}}$. In the chosen representation, the group $\mathbf{G_j^i}$ defined by Eq(2.4) is easily identified, and a set of planer surfaces is found $(m \neq 0)$, with l planes per row of each rep, where l and m are the usual orbital angular quantum numbers. The familiar molecular models in elementary chemistry and Physics texts show these nodes clearly.

For an N electron atom the situation is a bit more complicated, but nonetheless manageable. In this case the group $\mathbf{G}$ is the direct product of the permutation group for N particles with the diagonal subgroup of the rotation group in #n dimensions, The diagonal subgroup is isomorphic to $\mathbf{R^{3N}}$, and thus its reps are the same as those of $R^3$. The eigen subgroup is a direct product of the corresponding eigen subgroups for permutation and rotation. From the properties of direct product groups, the projection operator for $\mathbf{G}$ preserves the symmetries of each of its invariant subgroups. Eq (2.5) may be used separately for each of the invariant subgroups to locate the nodes, and the full invariant subgroups applied to find the nodal surfaces. In the process it is necessary to approximately define the projection operators
using group operations for which simultaneously rotate the coordinates of all electrons through the same angles. The rotational nodes are simply deduced from the rotation group in three dimensions, the permutation nodes are separately deduced.

Information about the permutation nodes contained in the Young's diagrams, which exibit one to one correspondence with the permutation group reps. What is true about these diagrams is that, separating space and spin, the variables in a given column must be mutually antisymmetric for that rep. Nothing else is implied about any other symmetries, as can be verified by using projection operators to describe symmetries of the reps. Next we deduce the nodal conditions for few atomic systems.

Let us consider the case of helium atom. Let us denote the wave function of such a system in spherical polar coordinates by $\psi(r_1, r_2, \vartheta_1, \vartheta_2, \varphi_1, \varphi_2)$.

For an S state the wave function is invariant to simultaneous rotation of the coordinates of electrons 1 and 2. Due to rotational invariance of an S state the above wave function can be described by $\psi(r_1, r_2, \vartheta)$ where $r_1$ is the distance of electron 1 from the origin, $r_2$ is the distance of electron 2 from the origin and $\vartheta$ is the polar angle between two electrons.

There are no nodes for the S rep, nodes are the consequences of permutation symmetries. Thus $^1S_0$ state is nodeless(because total spin is zero). The $^3S_1$ state is antisymmetric since total spin is odd(total spin one). To find the nodes for $^3S_1$ state, let us operate with the antisymmeric projection operator $P_-$ on $\psi$, giving

$P_- \psi(r_1, r_2, \vartheta) = \psi(r_1, r_2, \vartheta) - \psi(r_2, r_1, \vartheta)$

Obviously the above vanishes when $r_1 = r_2$.

In cylindrical coordinates the wave function can be described by $\psi(\rho_1, \rho_1, z_1, z_2, \varphi_1, \varphi_2)$. Since for two particle $P_z$ state the wave function is invariant under the simultaneous rotation of both the particles about the z axis, we can describe it by
$\psi(\rho_1, \rho_2, z_1, z_2, \cos(\varphi_1 - \varphi_2))$

Next we apply the symmetric projection operator to the above to get the nodal condition for $^1P_1$ state,

$P_+ \psi(\rho_1, \rho_2, z_1, z_2, \cos(\varphi_1 - \varphi 2)) = \psi(\rho_1, \rho_2, z_1, z_2, \cos(\varphi_1 - \varphi 2)) + \psi(\rho_2, \rho_1, \mathbf{z}_2, \mathbf{z}_1, \cos(\varphi_1 - \varphi 2))$

$P_+ \psi = 0$

$\Rightarrow \psi(\rho_1, \rho_2, z_1, z_2, \cos(\varphi_1 - \varphi 2)) = - \psi(\rho_2, \rho_1, \mathbf{z}_2, \mathbf{z}_1, \cos(\varphi_1 - \varphi 2))$

$\Rightarrow \psi(\rho_1, \rho_2, z_1, z_2, \cos(\varphi_1 - \varphi_2)) = \psi(\rho_2, \rho_1, -z_2, -z_1, \cos(\varphi_1 - \varphi_2))$ (Using symmetry property of P function) $\Rightarrow P_+\psi$ has zeros when

$z_1 = 0, z_2 = 0$ and

$z_1 = -z_2$ and $\rho_1 = \rho_2$

Similarly, $P_-\psi = (\rho_1, \rho_2, z_1, z_2, \cos(\varphi_1 - \varphi_2)) = 0$

$\Rightarrow z_1 = z_2$, and $\rho_1 = \rho_2$

Now we consider an S state of lithium. For 3 particle S state six coordinates will be sufficient to specify the configuration completely, i.e.,

$\psi(r_1, r_2, r_3, \cos\vartheta_{12}, \cos\vartheta_{23}, \cos\vartheta_{31})$

For the lithium $^2S_{1/2}$ state the row symmetrizer is defined by the Young's operator

$P' = e + (13)$

and column antisymmetrizer is defined by $Q' = e - (12)$

The complete projection operator is given by the Young's operator

$Y' = Q'P' = e - (12) + (13) - (132)$

Now $Y' \psi(r_1, r_2, r_3, \cos\vartheta_{12}, \cos\vartheta_{23}, \cos\vartheta_{31})$

$= \begin{matrix} \psi(r_1, r_2, r_3, \cos\vartheta_{12}, \cos\vartheta_{23}, \cos\vartheta_{31} \\ -\psi(r_2, r_1, r_3, \cos\vartheta_{21}, \cos\vartheta_{13}, \cos\vartheta_{23}) \\ +\psi(r_3, r_2, r_1, \cos\vartheta_{23}, \cos\vartheta_{12}, \cos\vartheta_{31}) \\ -\psi(r_3, r_1, r_2, \cos\vartheta_{13}, \cos\vartheta_{12}, \cos\vartheta_{23}) \end{matrix}$

$Y' \psi(r_1, r_2, r_3, \cos\vartheta_{12}, \cos\vartheta_{23}, \cos\vartheta_{31}) = 0$

$\Rightarrow r_1 = r_2$ and $\cos\vartheta_{13} = \cos\vartheta_{23}$ (2.8)

Similarly for the completely antisymmetric state,

We have $r_1 = r_2$ and $\cos\vartheta_{13} = \cos\vartheta_{23}$

or $r_2 = r_3$ and $\cos\vartheta_{21} = \cos\vartheta_{31}$

or $r_1 = r_3$ and $\cos\vartheta_{12} = \cos\vartheta_{32}$

Consider now an N electron atoms in which electrons 1 and 2 must be permutation antisymmeric; according to chosen Young's diagram. Then Eq(2.8) must be satisfied with $r_3$ replaced by $r_k$ $k = 3,....., N-2$. For every pair of permutation antisymmetric variables is repeated, as required by the corresponding Young's diagram. Thus. replacing (1,2,3) by (i,j,k) $N_p$ sets of N-2 simultaneous equations are obtained. The number of nodal surfaces becomes large, but the symmetry considerations and the group operations will in practice allow us to work in the interior of one bounded nodal surface.

In the usual approach to the many electron atom, Slater determinants of single particle functions are used initially. Even in the simplest case of helium atom first excited state, care must be taken to ensure that the proper nodal structure occurs with a single particle basis.
With a given constraint as to the form of the functions, a variational calculation may minimize the energy at the expense of improper nodes and symmetries. A more desirable approach of the would be to guarantee that the properties of the eigen and invariant subgroups are properly accounted for in the form of the trial function. A single Slater determinant cannot be expected to achieve this goal due to the presence of extraneous nodes not dictated by symmetry requirement, and thus leading to a higher energy than initially sought.

### 3.1 Analytical results: Nodal Equations for Simple Systems

We have shown the derivation nodal equations in the previous section. Let us now summarize all the results in this section.

For the ground state($^1S_0$), the wave function should be nodeless:

1. For the first excited state($^3S_1$) of helium, the nodal condition is $r_1 = r_2$

2. For a permutation symmetric $^1P_1$ state of helium. The nodal condition becomes

$z_1 = -z_2$ and $\rho_1 = \rho_2$ (notice that the condition $z_1 = z_2 = 0$ is included here.

This are not two independent conditions. If one combines them one would get a single equation

$\cos\theta_1 = -\cos\theta_2$ which will satisfy tiling property and divides the configuration space into nodal domains of constant signs '+' or '-'.

3 For the permutation antisymmetric $^3P_{2,1,0}$ state of helium, the nodal condition is

$z_1 = z_2$ and $\rho_1 = \rho_2$

Similarly these two equations can be combined into one single equation

$\cos\theta_1 = \cos\theta_2$

4 For the ground state of lithium, the conditions are

$^2S_{1/2}$, the conditions are $r_1 = r_2$ and $(\vec{r}_2 - \vec{r}_1)\cdot\vec{r}_3 = 0$

This equation also can be combined to give rise to

$\cos\theta_{23} = \cos\theta_{13}$

5 For the $^1P_{3/2,1/2}$ state of lithium, we get

$z_1 = z_2, r_1 = r_2$ and $(x_1 - x_2)x_3 + (y_1 - y_2)y_3 = 0$

The above equations except one in equation(5) are necessary and sufficient conditions for a wavefunction corresponding to a particular symmetry to be zero In each case(except the last one it is obvious that they are closed surfaces in their respective dimensional spaces, as previously discussed But in the case of equation (5) on the contrary, they form a lower dimensional subspace. So these equations do not form a closed surface. To perform a random walk for a physical system, we need to sample in a closed region of space as noted earlier. If we choose our trial functions which satisfy the conditions(5), additional assumptions are necessary so that the sampling can be done in a closed region of space. Unlike the fixed node approximation our nodal equations obey right symmetry condition required for the problem.

Numerical results: Wave function for $P_z$ state of Helium.

$$\psi(r_1, r_2, \vartheta_1, \vartheta_2) = (r_1 \cos\vartheta_1 + r_2 \cos\vartheta_2)[\exp(-\alpha_1 r_1 - \alpha_2 r_2) + \exp(-\alpha_2 r_1 - \alpha_1 r_2)].$$

It satisfies one of the two necessary nodal conditions for $P_z$ state, $z_1 = -z_2$ (proved in Sec 2.2)

Table 5
Numerical results and statistical data for the first excited state of He with trial function VII
$$\psi(r_1, r_2, \theta_1, \theta_2) = (r_1 \cos \theta_1 + r_2 \cos \theta_2)[\exp(-\alpha_1 r_1 - \alpha_2 r_2) + \exp(-\alpha_2 r_1 - \alpha_1 r_2)]$$
Table 5 shows the random walk calculation using n=900 as obtained from eqn(4). Fig 13 shows the plot of ln(zt)/t vs t including $1\sigma$ error bars for the uncertainty. In addition, a least square fit of the data starting at t=8 to the eqn(5) is shown in the figure 1.

| | | Scale =30,# of paths=600K | | | |
|---|---|---|---|---|---|
| t | zt | ln(zt) | ln(zt)/t | $\sigma$ | ln(zt)/t (ls fit) |
| 8 | 0.482733 | -0.728291 | -0.091036 | 0.000194 | -0.090174 |
| 16 | 0.802182 | -0.220419 | -0.013776 | 0.000382 | -0.014894 |
| 24 | 1.231027 | 0.207848 | 0.008660 | 0.000299 | 0.010199 |
| 32 | 2.333108 | 0.847201 | 0.026475 | 0.001216 | 0.022745 |
| 40 | 4.734821 | 1.554943 | 0.03873 | 0.001610 | 0.030273 |
| 48 | 6.348487 | 1.848216 | 0.038504 | 0.000930 | 0.035292 |

$$\lambda_0^{(0)} = -2.06460746 \quad \lambda_1 = -2.1250716(1)$$

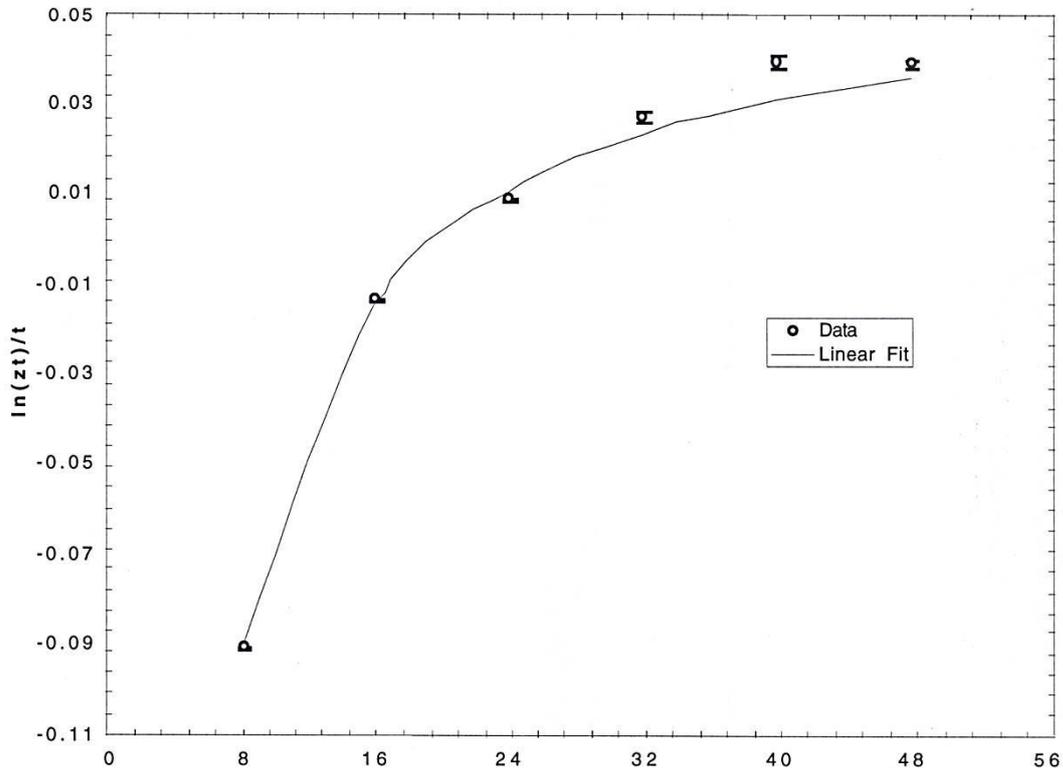

Figure 1  The plot of ln(zt)/t vs t for the wave function VII. The open circles are the numerically calculated values with $\pm 1\sigma$ error bars. The solid curve is a least square fit to eqn(5).

**5 Conclusions:**
In the usual approach to the many electron atom, Slater determinants of single particle functions are used initially. Even in the simplest case of helium atom first excited state, care must be taken to ensure that the proper nodal structure occurs with a single particle basis.
With a given constraint as to the form of the functions, a variational calculation may minimize the energy at the expense of improper nodes and symmetries. A more desirable approach of the would be to guarantee that the properties of the eigen and invariant subgroups are properly accounted for in the form of the trial function. A single Slater determinant cannot be expected to achieve this goal due to the presence of extraneous nodes not dictated by symmetry requirement, and thus leading to a higher energy than initially sought.

In conclusion, it has been demonstrated that it is possible to work out nodal structures and symmetries of eigenfunctions for some Hamiltonians. The eigengroup and invariant group of the row contain information about symmetries which are required for a given representation, and maybe employed to deduce the nodes without knowledge of the exact solution. The method works not only for the lowest eigenvalue of each rep, but can be generalized without difficulty to many electron atoms.